\documentclass[floatfix,aps,twocolumn,prb,amsmath,amssymb,superscriptaddress]{revtex4-1}
\newcommand{\Eu}{EuIr$_2$P$_2$}
\newcommand{\Eui}{Eu$^{+2}$}

\usepackage{lmodern}
\usepackage{longtable}
\usepackage{bm}
\usepackage[rgb]{xcolor}
\usepackage[author={Pablo S. Cornaglia}]{pdfcomment}
	\usepackage{multirow}
	\usepackage[pdftex]{graphicx}
	\usepackage[utf8]{inputenc}

	\usepackage{tabularx}
\usepackage{array} 
\usepackage[T1]{fontenc}
\newcolumntype{C}[1]{>{\centering\arraybackslash}p{#1}}

\graphicspath{{./figs/}}

\begin{document}
\title{Large magnetocaloric effect in the frustrated antiferromagnet EuIr$_2$P$_2$}
%\title{Magnetic properties of chiral \texorpdfstring{EuIr$_2$P$_2$}{EuIr2P2}}

\author{D. J. Garc\'{\i}a}
\affiliation{Centro At{\'o}mico Bariloche and Instituto Balseiro, CNEA, 8400 Bariloche, Argentina}
\affiliation{Consejo Nacional de Investigaciones Cient\'ificas y T\'ecnicas (CONICET), Argentina}
\author{D. G. Franco}
\affiliation{Centro At{\'o}mico Bariloche and Instituto Balseiro, CNEA, 8400 Bariloche, Argentina}
\affiliation{Consejo Nacional de Investigaciones Cient\'ificas y T\'ecnicas (CONICET), Argentina}

\author{Pablo S. Cornaglia}
\affiliation{Centro At{\'o}mico Bariloche and Instituto Balseiro, CNEA, 8400 Bariloche, Argentina}
\affiliation{Consejo Nacional de Investigaciones Cient\'ificas y T\'ecnicas (CONICET), Argentina}
\affiliation{Instituto de Nanociencia y Nanotecnolog\'{\i}a CNEA-CONICET, Argentina}

\begin{abstract}
  We present a theoretical analysis of the magnetocaloric effect in the frustrated antiferromagnet \Eu. Monte Carlo simulations indicate a large magnetic entropy change $\Delta S_M^{\text{max}} \sim  5.14$ J kg$^{-1}$K$^{-1}$ at $T=2.45$K for a moderate change in the external magnetic field $\Delta B=1.2$ Tesla. The highest value of $\Delta S_M$ seems to be associated with the presence of a multicritical point in the magnetic field versus temperature phase diagram (at $B_m\sim 1.2$ Tesla and $T_m\sim 2.2$ K) but it persists with similar large values for a broad range of temperatures $T\lesssim T_m$. Single crystal \Eu~magnetization vs magnetic field experiments allowed us to obtain the magnetocaloric effect for $T>2$ K. The results are consistent with the theoretical analysis and show a large magnetocaloric effect for $T\sim 2.5$ K. 
Finally, we present a phenomenological Landau functional analysis that provides a simplified description of the phase diagram and of the magnetocaloric effect. Although the magnetic frustration in this system leads to a rich phase diagram, we found no clear indication that it plays a significant role enhancing the magneticaloric effect, other than by reducing the transition temperatures and fields.

\end{abstract}


\maketitle

\section{Introduction}

Materials with magnetocaloric properties have attracted much attention over the years due to their use in solid state refrigeration technology\cite{zarkevich2020viable,tishin2016magnetocaloric}.
They offer an environmentally friendly and energy efficient alternative to conventional compression/expansion refrigeration\cite{gschneidner1999recent,pecharsky2006advanced}. The mechanism for magnetocaloric refrigeration is analogous to the one for conventional refrigeration, with the role of the external pressure being played by an external magnetic field. The production of a change $\Delta T_\text{ad}$ in the temperature of a material by applying an external magnetic field is known as the magnetocaloric effect (MCE). The temperature change $\Delta T_\text{ad}$ is determined by the change in the magnetic entropy $\Delta S_\text{M}(B)=S_\text{M}(B)-S_\text{M}(0)$ produced by the applied magnetic field $B$. 

A large magnetic entropy change $\Delta S_\text{M}(B)$ for a moderate change in the magnetic field $B$ (fields in the range $0$ to 2 Tesla allow the use of permanent magnets) is required for refrigeration applications.  $\Delta S_\text{M}(B)$ generally presents a peak as a function of the temperature which sets the temperature range of operation for a given material. A wide peak in $\Delta S_\text{M}(B)$ as a function of the temperature is therefore required for applications. The largest entropy changes have been reported in ferromagnetic materials, notably in Gd$_5$(Si$_2$Ge$_2$) and in MnAs$_{1-x}$Sb$_x$ associated with first-order phase transitions\cite{pecharsky1997giant,wada2001giant}.
Many other materials have been identified and used over the years for refrigeration at temperatures ranging from cryogenic to room temperature, but the quest for better materials is still ongoing \cite{LI2020153810,zarkevich2020viable}.

%A basic thermodynamic relation links the rate of entropy change with magnetic field at fixed temperature to the rate of change in the magnetization with temperature at fixed magnetic field
%$\left( \frac{\partial S}{\partial B} \right)_T=\left( \frac{\partial M}{\partial T} \right)_B$. 
%Large values of $\frac{\partial M}{\partial T}$ are expected at the vicinity of ferromagnetic-paramagnetic second-order phase transitions. 
Although antiferromagnetic (AFM) materials have drawn relatively less attention than the ferromagnetic ones, they can also show a large MCE\cite{samanta2007giant,das2018giant}. The external magnetic field can induce a metamagnetic transition in which the AFM order is suppressed leading to a large entropy change. Magnetic frustration effects in these materials can play an important role for the MCE. On the one hand, they can lead to a considerable reduction of the critical magnetic field needed to produce a metamagnetic transition. On the other hand, frustration can lead to highly degenerate states and, it has been argued\cite{zhitomirsky2003enhanced,zhitomirsky2004magnetocaloric,sosin2005magnetocaloric,HONECKER20061098}, to an enhancement of the MCE. 

%In particular  ErRu$_2$Si$_2$ compound, which presents a first-order magnetic field-induced AFM to FM transition.

The so-called giant MCE has been reported in frustrated antiferromagnets \cite{midya2012giant,das2018giant} but the role of frustration remains unclear. This calls for the exploration of the MCE in other frustrated systems, and for more theoretical analyses to help a better understanding of the role of frustration in the MCE in particular, and the conditions to optimize the MCE in general\cite{biswas2019designed,ivchenko2021emerging}. In this article we focus on the frustrated antiferromagnet \Eu\ \cite{lux1993kristallstrukturen}. The magnetic frustration in this material is due to both competing interactions
%(the possitive Curie-Weiss temperature indicates predominant ferromagnetic interactions) 
and geometric frustration due to its triangular structure. The large magnetic moments ($J=7/2$) at the \Eui\ ions of this compound dominate the magnetic properties of this system. \Eu~presents two phase transitions as a function of the temperature at low magnetic fields: a paramagnetic to collinear AFM transition and a collinear to non-collinear antiferromagnetic-antiferromagnetic transition \cite{franco2021synthesis,vildosola2021magnetic}. It also presents a metamagnetic (spin-flop) transition as a function of the magnetic field. \Eu's rich phase diagram makes it an interesting playground to analyze the role of frustration in the MCE.

We analyze the MCE in \Eu\ using both classical Monte Carlo and a Landau functional. We find a wide range of temperatures with a large $\Delta S_\text{M}(B)$ for moderate magnetic fields (lower than 2 Tesla), which is a desirable property for cryogenic applications. The Landau functional analysis suggests that the system presents a multicritical point where three second-order and one first-order phase transition lines meet. It is in the vicinity of this point that the largest magnetic entropy changes are obtained. We also report magnetization measurements in \Eu~ single crystals as a function of the temperature ($T>2$ K) and the external magnetic field. The MCE obtained from these measurements is in agreement with the theoretical analysis.

The rest of this paper is organized as follows: 
In Sec. \ref{sec:model} we present the Hamiltonian for the magnetic moments of \Eu\ and the Monte Carlo magnetic phase diagram. 
In Sec. \ref{sec:landau} we introduce the Landau functional used to describe the magnetic phase diagram in a phenomenological way, and compare with the Monte Carlo results.
In Sec. \ref{sec:MCE} we present the results for the MCE obtained using both Monte Carlo simulations and the Landau functional approach.
In Sec. \ref{sec:EMCE} we present the results for the MCE  obtained using magnetization measurements of \Eu~single crystals.
Finally, in Sec. \ref{sec:concl} we summarize our main results and conclusions.

%
%Hexagonal lattice (frustration + disorder) \cite{pakhira2016large}
%Use of the MC effect to analyze phase transitions \cite{chakraborty2019magnetocaloric}

%Usually, the study of MCE is focused on the ferromagnetic materials, such as Gd5Si2Ge2,5 MnFeP0.45As0.55,6 MnAs,7 ErCo2,8 HoCo2,9 and DyCo2.10 These materials display large MCE due to the first-order magnetic disorder-to-order transition from paramagnetic to ferromagnetic state.
%However, a little attention has been explored on antiferromagnetic systems that can also exhibit large MCE because of the first-order magnetic field-induced order-to-order phase transition from antiferromagnetic (AFM) to ferromagnetic (FM) state.13 In this context, we have studied the MCE of antiferromagnetic ErRu2Si2 compound, which shows the evidence of first-order magnetic field-induced order-to-order phase transition. AFM to FM.

%Insulating materials desired for MC applications???. Large thermal conductivity is needed\ldots

%Low hysteresis too. \textcolor{red}{preguntar a Diego por el \Eu}.

%What happens in a polycrystal?

%\section{Model for the magnetic interactions} 
\section{Magnetic Hamiltonian and Monte Carlo results}\label{sec:model}
A detailed analysis of the magnetic phase diagram of \Eu\ was provided in Ref. \onlinecite{vildosola2021magnetic}. In this Section we present a brief summary of the main results. 
\subsection{Crystal structure of the \Eui\ ions}
The magnetism in \Eu\ is dominated by the magnetic moments in the \Eui\ ions. Hund's rules indicate a $J=7/2$ angular momentum which is consistent with magnetic susceptibility measurements\cite{franco2021synthesis}.
The unit cell of \Eu\ contains three Eu atoms (Eu$_1$, Eu$_2$, and Eu$_3$). 
The Eu$_2$ ions form a hexagonal lattice that can be parametrized with the primitive vectors $\mathbf{a}=(a,0,0)$, $\mathbf{b}=(-a/2,\sqrt{3}a/2,0)$, and $\mathbf{c}=(0,0,c)$.
The Eu$_1$ ions form a hexagonal lattice shifted by $\delta \mathbf{r}_1 = (-0.094 a,  0.523 a,- c/3)$ from the Eu$_2$ lattice, while the hexagonal lattice of Eu$_3$ ions is shifted by $\delta \mathbf{r}_3 = (a/2, 0.18 a, c/3)$ from the Eu$_2$ lattice (see Fig. \ref{fig:Eucrystal}). 
The arrangement of the \Eui\ ions in the lattice can be viewed as a stacking of hexagonal layers along the $\hat{z}$-axis (similar to the ABC stacking of the face centered cubic lattice).

\subsection{Magnetic Hamiltonian}
\begin{figure}[t]
    \begin{center}
       \includegraphics[width=0.3\textwidth]{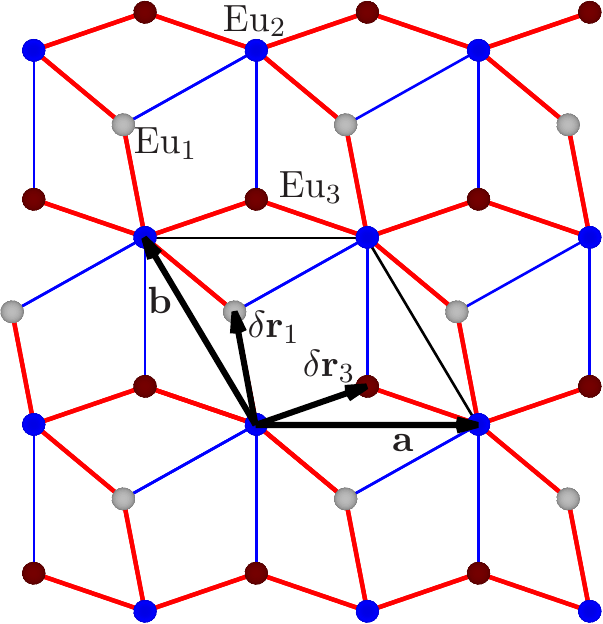}
    \end{center}
    \caption{(Color online) Top view of the EuIr$_2$P$_2$ crystal structure showing only the \Eui\ ions.
  %    First (second) nearest neighbours are connected by red (blue) lines. 
      The three \Eui\ ions in the unit cell (Eu$_1$, Eu$_2$, and Eu$_3$ are represented using different colors) form hexagonal lattices in the basal plane. The first (second) nearest neighbours of the Eu$_2$ atoms (blue symbols) are indicated using thick red (thin blue) solid lines.
 }
    \label{fig:Eucrystal}
\end{figure}

The magnetic properties of this compound are well described by an exchange coupling Hamiltonian including a magnetic anisotropy term \cite{vildosola2021magnetic}:
\begin{align}
	\mathcal{H}_{m}&= -\sum_{\left< i,j\right>} J_{1} \mathcal{J}_i \cdot {\mathcal J}_j \nonumber  - \sum_{\left<\left< i,j\right>\right>} J_{2} \mathcal{J}_i\cdot \mathcal{J}_j \\& -D\sum_i (\mathcal{J}_{i}\cdot \hat{z})^2 -g\mu_B\sum_i {\mathcal{J}_i}\cdot {\bf B},
    \label{eq:magham}
\end{align}
with the coupling constants ($J_1$, $J_2$), and the magnetic anisotropy coefficient $D$ estimated through {\it ab initio} calculations of the total energy of the system \cite{vildosola2021magnetic}. The exchange coupling between nearest neighbour (nn) magnetic moments $J_{1}\sim0.28$K is ferromagnetic, the next-nearest-neighbour (nnn) exchange interaction $J_2\sim -0.45$K is antiferromagnetic and the anisotropy $D\sim0.2$K is of the easy axis type. An \Eui\ ion at position $i$ has a magnetic moment $\mathcal{J}_i$, four nn and two nnn (see Fig. \ref{fig:Eucrystal}). The Hamiltonian also contains a Zeeman coupling of the magnetic moments to an external magnetic field  ${\bf{B}}$. 
In what follows we use units such that $k_B$=1 and we focus on the case of a magnetic field parallel to the $\hat{z}$-axis. 

\subsection{Phase diagram}
\begin{figure}[t]
    \begin{center}
       \includegraphics[width=0.475\textwidth]{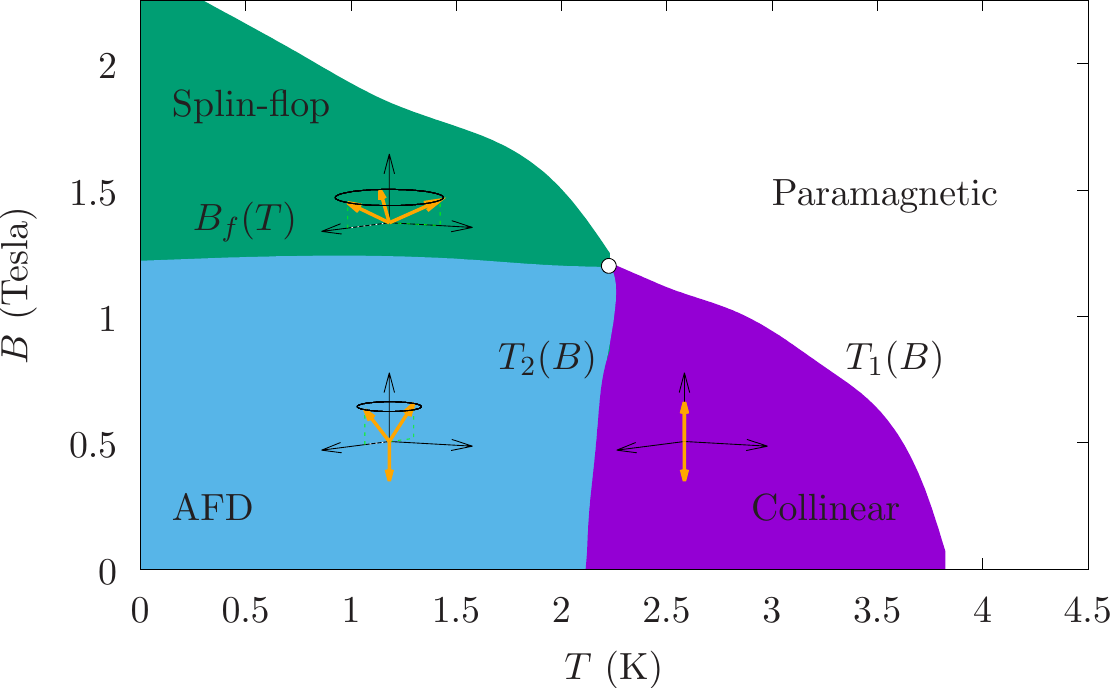}
    \end{center}
    \caption{Magnetic phase diagram for an external magnetic field parallel to the easy axis. Four phases can be identified: paramagnetic for high temperatures, collinear antiferromagnet, non-collinear antiferromagnet (AFD) and spin-flop state for low temperatures and intermediate fields (see Ref. [\onlinecite{vildosola2021magnetic}]). The white disk marks the position of a multicritical point.
    %The points were anomalies in the specific heat and or the magnetization.
    }
    \label{fig:PD_MC}
\end{figure}

The phase diagram of Fig. \ref{fig:PD_MC} was constructed through classical Monte Carlo numerical simulations of the magnetic Hamiltonian, considering the large $J=7/2$ magnetic moments of the \Eui\ ions as classical\cite{bauer2011alps,ALBUQUERQUE20071187,garcia2020magnetic,vildosola2021magnetic}.
It presents four phases which result from the competition between the exchange interactions and the easy-axis anisotropy. At zero field, there are two phase transitions: a high temperature transition from a paramagnet to a collinear antiferromagnet at $T_1$, followed by a transition to the non-collinear antiferromagnet phase  (that we call AFD) at a lower temperature $T_2$. The collinear antiferromagnet is stabilized by its high configurational degeneracy\cite{vildosola2021magnetic}. Increasing the magnetic field from the AFD phase, a spin-flop (s-f) transition is observed which is accompanied by a jump in the magnetization. 

The transition lines $T_1(B)$ and $T_2(B)$ and the spin-flop field  $B_f(T)$ meet (within the numerical accuracy of our Monte Carlo calculations) at a multicritical point. While the spin-flop transition is clearly of the first-order type, $T_1(B)$ and $T_2(B)$ and the spin-flop to paramagnetic transitions seem to be continuous.  
%Although the transition temperatures $T_1$ and $T_2$ are underestimated in this analysis, the critical field for the spin-flop transition $B_f$ and the crossing of $T_1(B)$ and $T_2(B)$ are in very good quantitative agreement with the reported experimental results \cite{franco2021synthesis,vildosola2021magnetic}.

A phase diagram in which three second-order transitions lines and a first-order transition merge into a multicritical point is consistent with general thermodynamic principles \cite{leggett1974implications,yip1991thermodynamic}. As we show below, the phase diagram of Fig. \ref{fig:PD_MC}  and the corresponding phases can be described qualitatively using a Landau functional.

\section{Landau functional analysis} \label{sec:landau}
Landau functionals have proven very useful to analyze and interpret MCE experiments \cite{yamada2003itinerant,amaral2004magnetoelastic}.
We follow Refs. \cite{zhu1987phenomenological,plumer1988multicritical,quirion2011magnetoelastic}
and build a Landau functional to provide a phenomenological description of the phase diagram presented in Fig. \ref{fig:PD_MC}.

\begin{figure}[t]
    \begin{center}
       \includegraphics[width=0.4\textwidth]{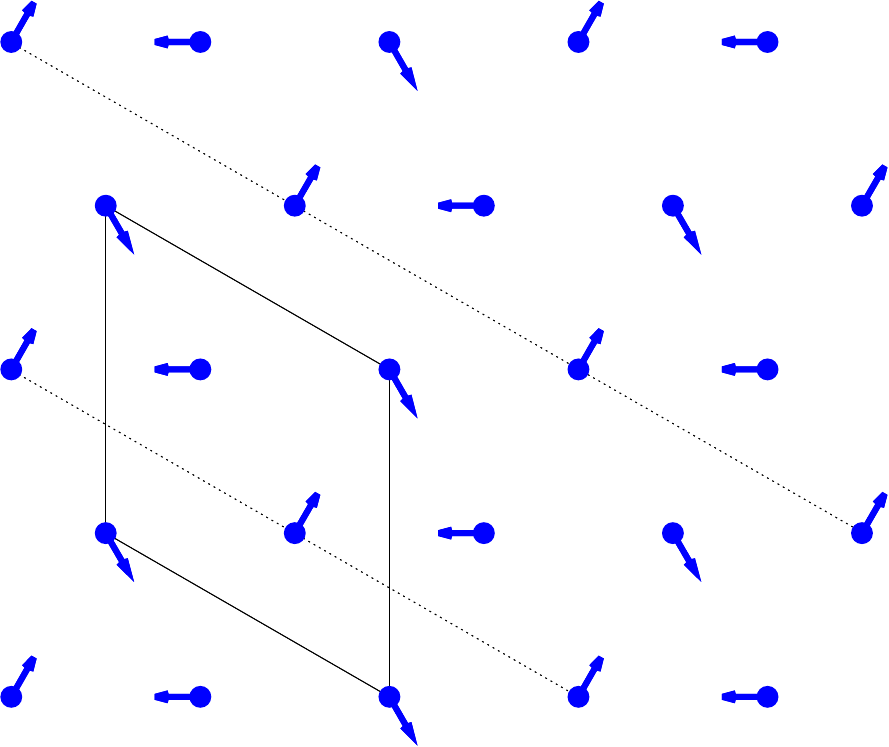}
    \end{center}
    \caption{ Schematic representation of the magnetic moments in the 120$^\circ$ configuration for a layer of Eu$_1$ atoms. The dotted lines indicate chains of parallel magnetic moments. A single wave vector $\mathbf{Q}$ perpendicular to these lines and of a length of $|\mathbf{Q}|=2\pi/d$, where $d$ is the distance between dotted lines, is enough to characterize the magnetic moment configuration. The solid lines indicate the shape of the magnetic unit cell in the $a$-$b$ plane.
    %The points were anomalies in the specific heat and or the magnetization.
    }
    \label{fig:120}
\end{figure}

We focus our analysis on the case where the magnetic field is along the easy axis $\mathbf{B}\parallel \hat{z}$.
The magnetic configurations obtained from the Monte Carlo analysis can be parametrized using a complex order parameter vector
\begin{equation}\label{eq:op}
  \mathbf{s}=s \cos \beta \left[\sin\theta\, \hat{n}_1 + \cos \theta\, \hat{z}\right]+i s \sin\beta \,\hat{n}_2
\end{equation} 
and a single wave vector $\mathbf{Q}=2 \pi(\frac{1}{3a},\frac{1}{\sqrt{3}a}, \frac{1}{2 c})$
where $a$ is the lattice parameter of the hexagonal layer, $c$ the distance between layers of the same type of Eu, and $\hat{n}_1$ and $\hat{n}_2$ are two orthogonal unit vectors in the basal plane. The Landau functional has to be independent of the choice of $\hat{n}_1$ and $\hat{n}_2$ due to the rotational invariance of the magnetic Hamiltonian around the $\hat{z}$-axis.

%$\mathbf{Q}=\frac{\mathbf{b}_1 +\mathbf{b}_2}{3} +\mathbf{b}_3/2$
%where $\mathbf{b}_i$ are the primitive reciprocal vectors of the Eu1 lattice.

The magnetic moment on an Eu$_i$ ion ($i=1,\,2,\,3$) at the unit cell associated with the Bravais lattice vector $\mathbf{R}$ can be written as the sum of a ferromagnetic (uniform magnetization $\mathbf{m}$) and an antiferromagnetic component.  
\begin{equation}
  \mathcal{S}_i[\mathbf{R}]=m\hat{z}+ \mathbf{s}e^{i(\mathbf{Q}\cdot\mathbf{R}+ \delta_i)}+ \mathbf{s}^*e^{-i(\mathbf{Q}\cdot\mathbf{R}+\delta_i)}
  \label{eq:parametrization}
\end{equation}
where $\delta_1=-\pi/3$, $\delta_2=0$, and $\delta_3=\pi/3$.

%The configuration of other Eui magnetic moments only differs in a phase $\delta_i$ \textcolor{red}{ver cómo}

Setting $\theta=\pi/2$ and $\beta=\pi/4$ in Eq. (\ref{eq:op}) results in the well known 120$^\circ$ structure for classical spins on a hexagonal lattice\cite{diep2013frustrated} with the spins contained in the basal plane (i.e. no projection along $\hat{z}$-axis) as schematized in Fig. \ref{fig:120}. Adding a finite magnetization $m$ to this configuration leads to the spin-flop state. For $\theta=0$ and $\beta=0$ the spins are along the $\hat{z}$-axis which results in a collinear antiferromagnet. Finally, $\theta=0$ and $0< \beta<\pi/4$ corresponds to the AFD configuration. In this latter case, the projection along the $\hat{z}$-axis is larger than the projection on the basal plane due to the easy axis anisotropy. The paramagnetic case corresponds to $s=0$ but there can be a finite magnetization $m\neq0$ induced by the external magnetic field.

Keeping terms allowed by symmetry up to fourth order in $\mathbf{s}$ and $m$ leads to a functional which is convenient to write in terms of $\zeta=\cos\theta$, $s=|\mathbf{S}|$, $s_p=s\sin\beta$ and $m$ (see Refs. \onlinecite{zhu1987phenomenological,plumer1988multicritical,quirion2011magnetoelastic}):
\begin{align}
  \mathcal{F}(s,\zeta^2,s_p,m)&=\mathcal{F}_0  + \frac{A_0 m^2}{2}+\frac{B_3 m^4}{4}-B \text{m} \nonumber \\ 
 &+ A_Q s^2+\frac{B_1 s^4}{2}+\frac{B_6 B^2 s^2 }{2} \label{eq:flandau} \\
 &+2 B_2  s_{p}^2 \left(s_{p}^2-s^2\right)-A_D \zeta^2 (s^2-s_p^2)\nonumber\\ 
 &+2 B_4 \zeta ^2 m^2 \left(s^2-s_{p}^2\right)+B_5 m^2 s^2.\nonumber 
  \end{align}
The basal plane component $s_\perp$ of the order parameter can be written in terms of $s$, $s_p$, and $\zeta$ as $s_\perp=\sqrt{(s^2-s_p^2)(1-\zeta^2)+s_p^2}$, while its $\hat{z}$-axis component is given by $s_z= \sqrt{\zeta^2(s^2-s_p^2)}$.

The easy axis anisotropy is taken into account in $\mathcal{F}$ by the term $-A_D s_z^2\equiv -A_D \zeta^2(s^2-s_p^2)$, with $A_D>0$. 
We include a temperature dependence in the parameter $A_Q=A(T-T_Q)$ which leads to the usual Landau second-order phase transition to an antiferromagnet ($s\neq 0$) at $T_Q$ in the absence of magnetic anisotropy and for $B=0$. We also set $A_0=A (T-T_0)$ to obtain a Curie-Weiss behavior of the magnetic susceptibility.  The other parameters were chosen (see Appendix \ref{ap:landau}) to reproduce the qualitative behavior of the Monte Carlo phase diagram of Fig. \ref{fig:PD_MC}.
\begin{figure}[t]
    \begin{center}
       \includegraphics[width=0.45\textwidth]{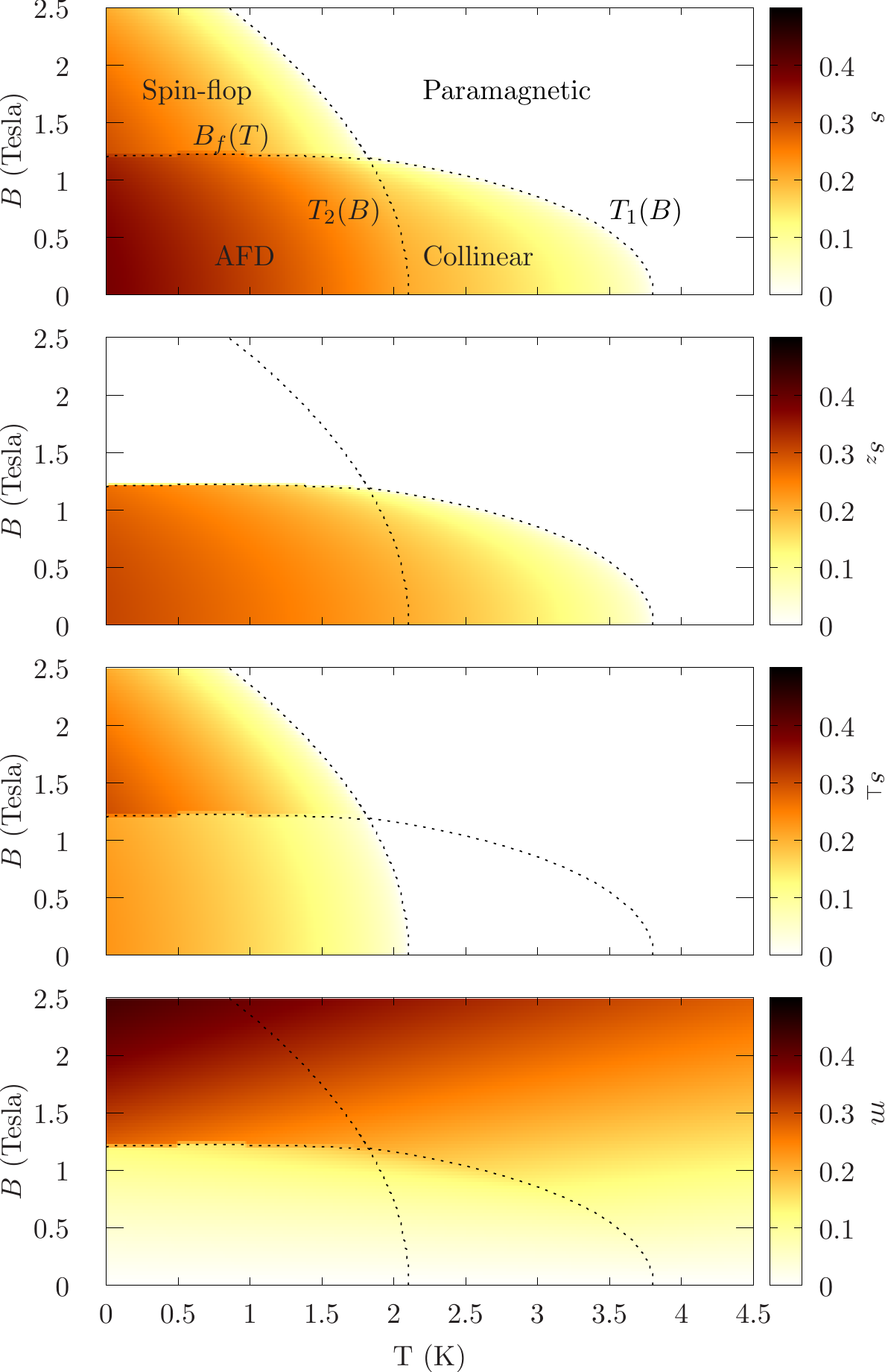}
    \end{center}
    \caption{Modulus $s$ of the order parameter, its components $s_z$, $s_\perp$, and the magnetization $m$ as a function of the temperature and the external magnetic field. The dotted lines indicate the different phase transitions.}
    \label{fig:OPlandau}
\end{figure}

The phase diagram of Fig. \ref{fig:OPlandau} is obtained minimizing the Landau functional $\mathcal{F}$ w.r.t. $s^2$, $\zeta^2$, $s_p$, and $m$ for different values of the temperature and the external magnetic field. The modulus $s$ of the order parameter, its components  $s_z$,  $s_\perp$, and the magnetization $m$ are also presented in the figure.
The order parameter $s$ is nonzero in all the ordered phases and vanishes continuously at the phase transition lines to the paramagnetic state.

At zero external magnetic field $B=0$, the magnetization vanishes ($m=0$) and there are two continuous phase transitions at temperatures $T_1= T_Q+\frac{A_D}{A}$ and $T_2= T_1-\frac{B_1}{2B_2}\frac{A_D}{A}$. $T_1$ marks a continuous transition from paramagnet ($s=0$) to collinear antiferromagnet ($s=s_z\neq 0$, $s_\perp$=0) where the basal plane components of the spin density vanish. At $T=T_2$ there is a second-order transition from collinear AFM to the AFD state ($s\neq 0$, $s_\perp\neq 0$, $s_z\neq 0$).
The difference $T_1-T_2=\frac{A_D}{A}\left( 1+\frac{B_1}{2 B_2} \right)$ is proportional to the anisotropy parameter $A_D$ which is consistent with the Monte Carlo results \cite{vildosola2021magnetic}. 

Turning on an external magnetic field the temperature difference $|T_1(B)-T_2(B)|$ is reduced and the two transitions merge into a single point at high enough fields. At the same point merge two additional transition lines: a first-order transition from the AFD state to the spin-flop state ($s_z =0$, $s=s_\perp\neq0$) where the antiferromagnetic order is a 120$^\circ$ structure with the magnetic moments contained in the basal plane, and a second-order transition line separates the spin-flop state from the paramagnet.
The spin-flop transition at field $B_f(T)$ is accompanied by a jump in $m$. The jump decreases its height as the temperature is increased and vanishes at the multicritical point as observed in the Monte Carlo simulations\cite{vildosola2021magnetic}.

%The slope of the flop transition line is governed by the parameter $B_5$.

\subsection{Comparison to an order parameter analysis of the Monte Carlo results}

To make a more detailed comparison between the Monte Carlo and the Landau functional results we calculate the structure factor:
\begin{equation}
  \mathbf{S}(\mathbf{q})=\frac{1}{J N}\sum_{\ell}\mathcal{J}_\ell e^{ i \mathbf{q}\cdot \mathbf{R}_\ell},
  \label{eq:strfac}
\end{equation}
where $N$ is the number of \Eui\ ions used in the Monte Carlo simulations, $\mathbf{R}_\ell$ is the position of the magnetic moment $\mathcal{J}_\ell$, and $J=7/2$ is the magnetic moment at the \Eui~ions.
 \begin{figure}[t]
\begin{center}
       \includegraphics[width=0.45\textwidth]{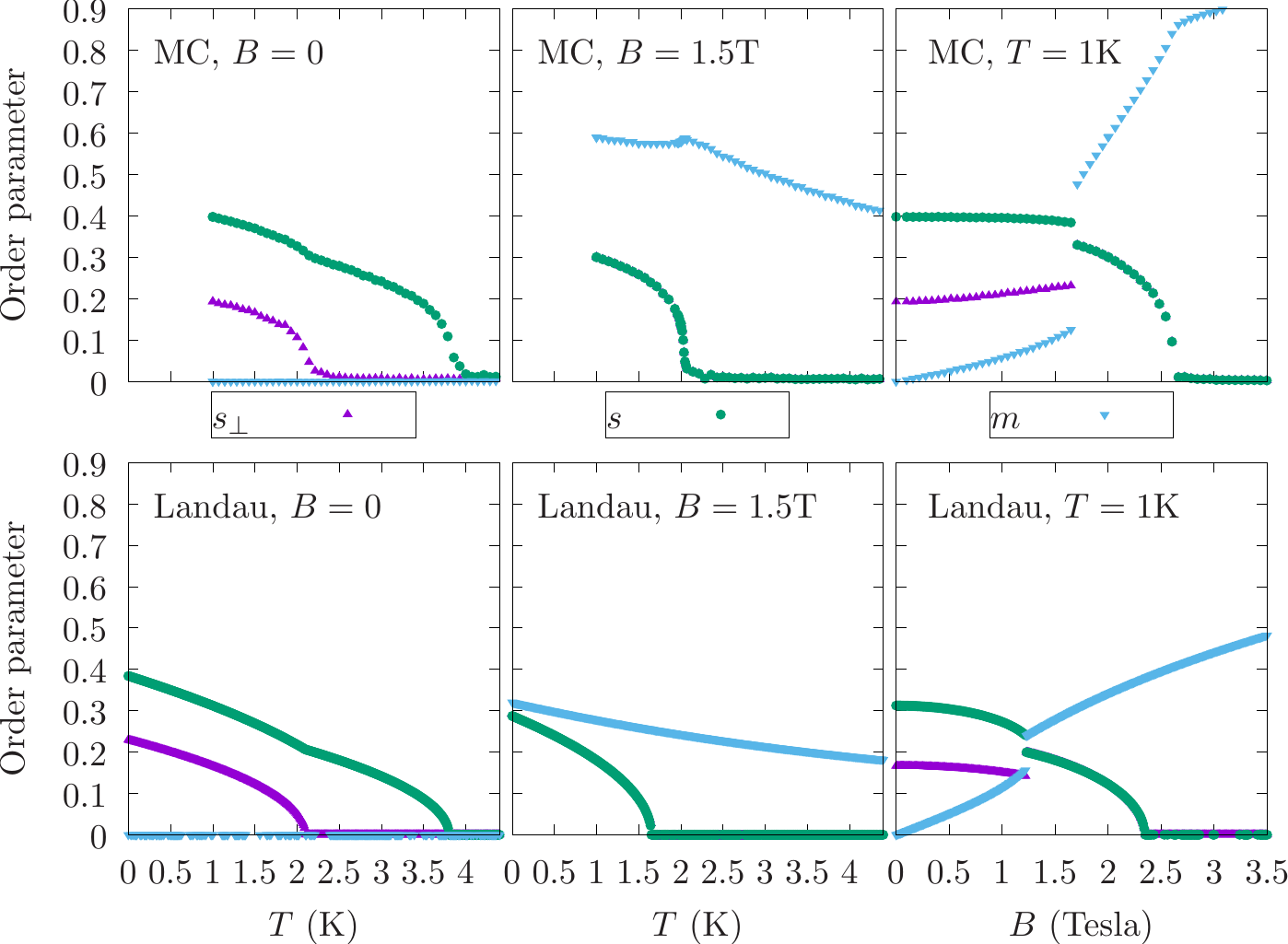}
    \end{center}
    \caption{(Color online) Comparison of the order parameters obtained from the Monte Carlo simulations (top panels) with the ones from the Landau functional (bottom panels) for different magnetic field and temperature cuts of the phase diagram (note that $s=s_\perp$ for $B=1.5$ T in the whole range of temperatures).}
       \label{fig:compOP}
\end{figure}

For a spin configuration that can be described using Eq. (\ref{eq:parametrization}) we only expect non-zero values of $\mathbf{S}(\mathbf{q})$, for $\mathbf{q}=0$, $\mathbf{q} =\mathbf{Q}$, and symmetry related wave vectors. This is precisely what we obtain in our Monte Carlo simulations. The structure factor allows us to calculate the order parameter vector $\mathbf{s}$. For the comparison with the Landau functional results we calculate $s^{MC}=|\mathbf{S}(\mathbf{Q})|$, $s^{MC}_\perp=\sqrt{|\mathbf{S}(\mathbf{Q})|^2-|\hat{c}\cdot \mathbf{S}(\mathbf{Q})|^2}$, and $m^{MC}=|\hat{c}\cdot\mathbf{S}(0)|$. 

Figure \ref{fig:compOP} presents the results for the order parameters obtained from the Monte Carlo (top panels) and the Landau functional (bottom panel) calculations for three paths in the phase diagram that go across the different phase transitions.  The Monte Carlo results were obtained in systems of $24\times24\times24$ unit cells averaging $|\mathbf{S}(\mathbf{q})|^2$ and $|\hat{z}\cdot\mathbf{S}(\mathbf{q})|^2 $, over 4 configurations of the magnetic moments, separated by $10^5$ MC steps, for each temperature and magnetic field.
The behavior of the order parameters presents a qualitative agreement between the two methods, which indicates that the Landau functional captures the main physics of the magnetic phase diagram.

\section{Magnetocaloric effect} \label{sec:MCE}
We calculated the magnetic entropy change $\Delta S_\text{M}$ from the classical Monte Carlo results for the specific heat and the magnetization. The integration of the magnetic contribution to the specific heat $C_B$ to obtain the magnetic entropy,
\begin{equation}
	S_\text{M}(T)= \int_0^T \frac{C_B}{T} d T,
\end{equation}
 has the drawback that classical spins have a finite specific heat as $T\to 0$ which leads to a spuriously divergent entropy. Since we are interested in entropy differences, this divergence can be cured setting a high temperature reference entropy (see e.g. Ref. \onlinecite{garcia2020magnetic}). %The calculation of the entropy change at a first-order transition using the specific heat is however numerically costly. 

 The isothermal entropy change for a reversible change in the magnetization can also be obtained integrating a Maxwell relation. As proposed by Amaral {\it et al.}, to correctly obtain the change in the entropy at a first-order phase transition it is convenient to use the following expression:\cite{amaral2010estimating,mashirov2017revision}
\begin{equation}
	\Delta S_M(T,B)=\frac{\partial}{\partial T} \int_0^{B} M(T,B^\prime) dB^\prime.
  \label{eq:maxwell}
\end{equation}
%in which the order of the derivative and the integral in the right hand side has been changed.
%For this expression to be valid the magnetization needs to be obtained at equilibrium

Another useful relation to analyze the MCE is the magnetic Clausius-Clapeyron relation which links the slope of the spin-flop first-order transition line to the magnetization  and entropy jumps across the transition, ($\Delta M=M_{AFD}-M_{s\text{--}f}$ and $\Delta S_M=S_{M}^{AFD}-S_{M}^{s\text{--}f}$, respectively):
\begin{equation}
	\frac{dB_f}{dT}=-\frac{\Delta S_M}{\Delta M},
  \label{eq:CC}
\end{equation}
\begin{figure}[t]
    \begin{center}
       \includegraphics[width=0.4\textwidth]{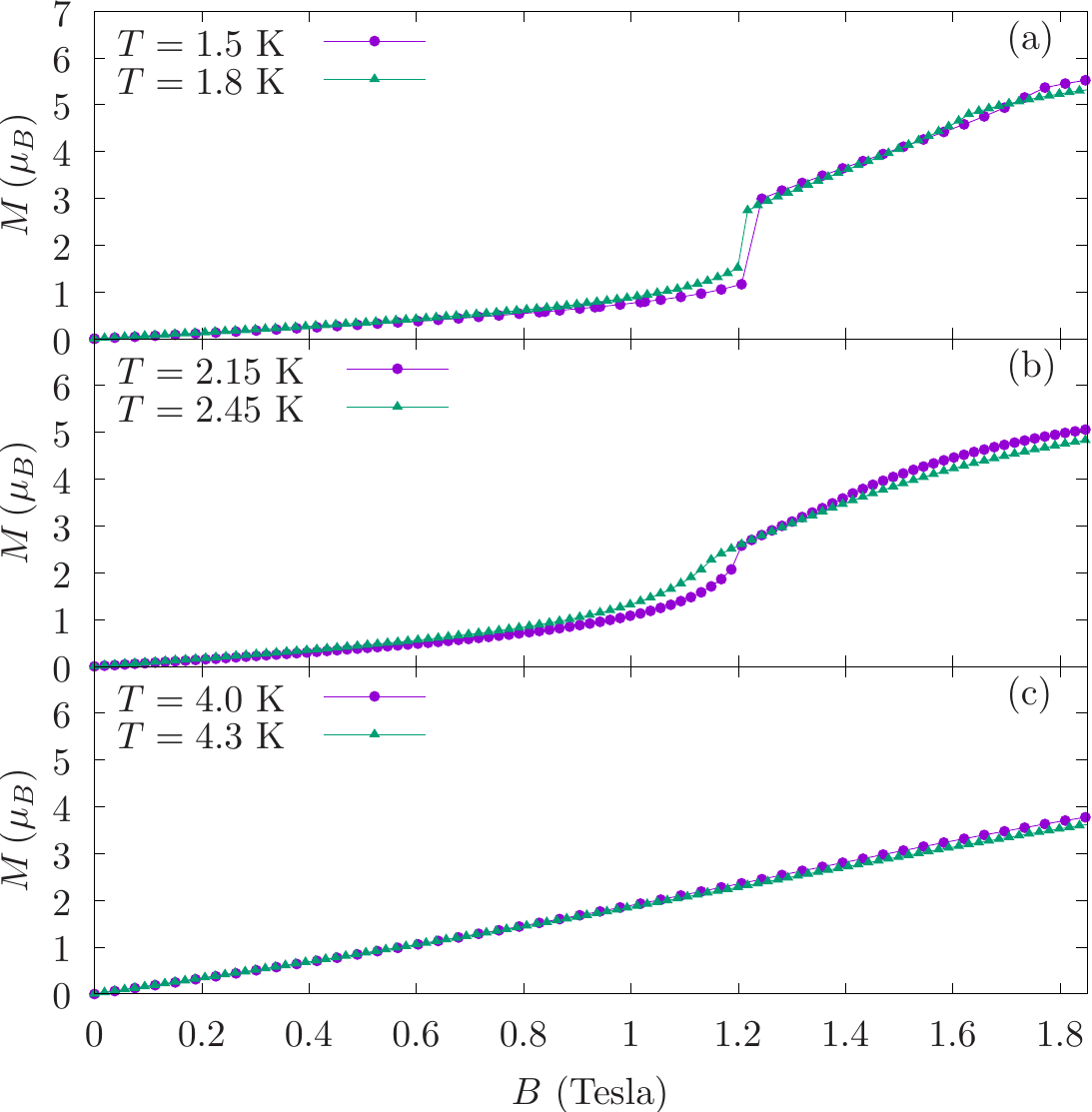}
    \end{center}
    \caption{ Magnetization along the $\hat{z}$-axis, per \Eui\ ion, as a function of the intensity of the  external magnetic field $B \hat{z}$. (a) Low temperature regime showing a jump in the magnetization associated with the spin-flop transition and a kink at higher fields associated with the spin-flop to paramagnetic transition. (b) Intermediate temperature regime: the larger changes in the magnetization for a given change in the temperature are obtained for fields near the collinear AFM to paramagnet transition line. (c) High temperature regime: the system is in the paramagnetic phase for all values of $B$.   }
    \label{fig:MagCMC}
\end{figure}

Figure \ref{fig:MagCMC} presents the Monte Carlo results for the magnetization per \Eui\ ion as a function of the magnetic field for three different temperature regimes. At low temperatures there is a clear jump in the magnetization at $B\sim 1.2$ Tesla which is associated with the spin-flop transition. An increase in the temperature in this regime produces a reduction in the magnetization jump which is dominated by an increase in the magnetization for fields below the spin-flop transition field. In the intermediate regime (middle panel in Fig. \ref{fig:MagCMC}) there is a kink in the magnetization at the collinear AFM to paramagnetic transition line. The largest changes in the magnetization for a given change in the temperature occur close to this kink. Finally, in the high temperature regime, the system is in the paramagnetic phase and an increase in the temperature produces a decrease in the magnetization.

\begin{figure}[t]
    \begin{center}
       \includegraphics[width=0.45\textwidth]{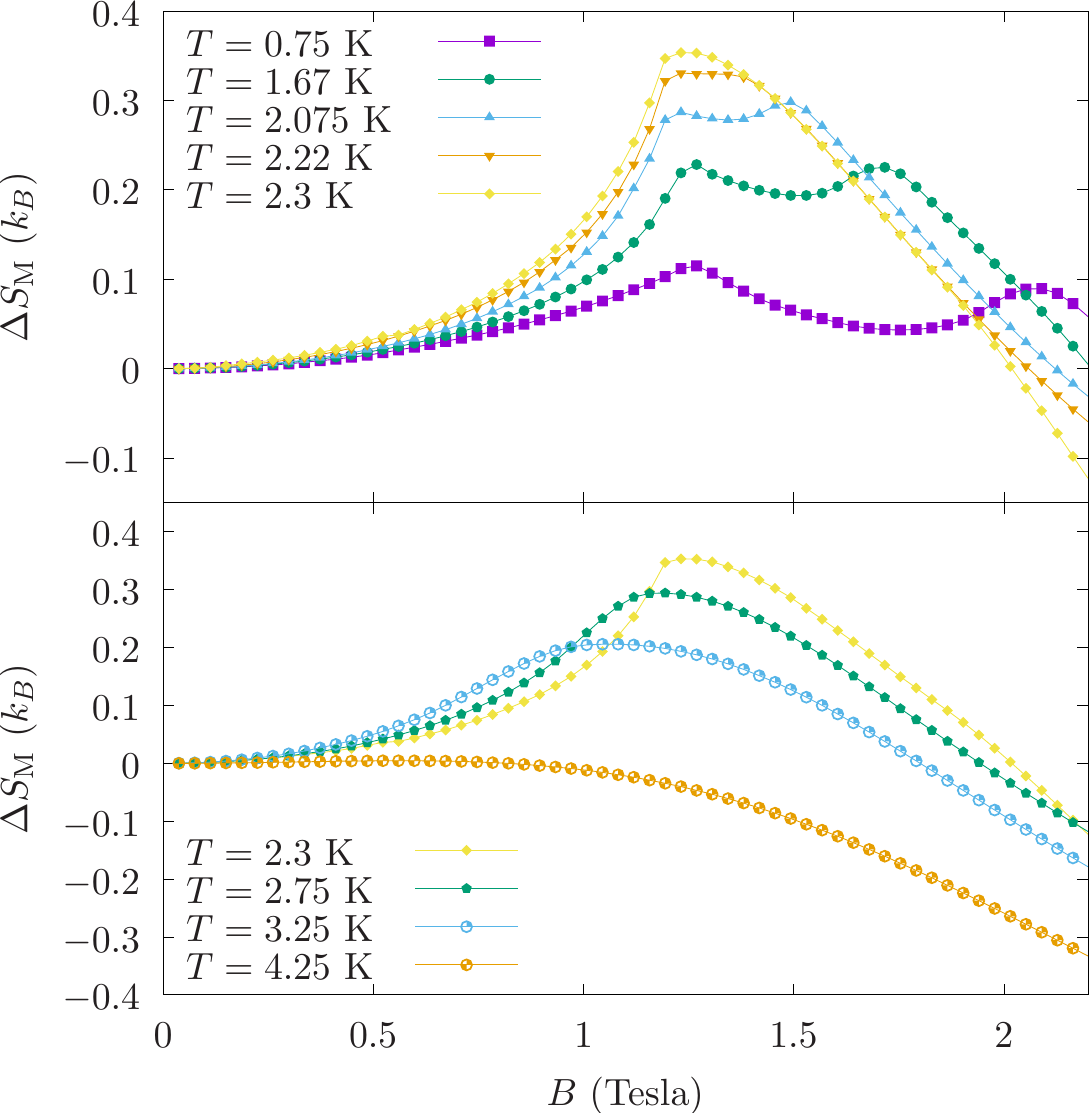}
    \end{center}
    \caption{Magnetic entropy change $\Delta S_\text{M}$ per \Eui\ ion calculated using a Monte Carlo approach for different values of the temperature and the final magnetic field.}
    \label{fig:MCCMC}
\end{figure}

Figure \ref{fig:MCCMC} presents the MCE, calculated using Eq. (\ref{eq:maxwell}) as a function of the magnetic field at different temperatures. 
In the lower temperature regime two peaks are obtained in $\Delta S_M$ as a function of the external field. The one at the higher field is associated with the spin-flop to paramagnetic transition and the other one is due to the spin-flop transition. A jump in $\Delta S_M$ is expected at the spin-flop transition. However, since the slope of $B_f(T)$ is very small (the temperature dependence is smaller than the numerical precision used in the determination of the transition line), the jump in the entropy is also expected to be small [see Eq. (\ref{eq:CC})]. As the temperature is increased the two peaks become more intense, the lower field peak remains at nearly the same field, while the position of the other peak shifts to lower fields following the spin-flop to paramagnetic phase transition line. The highest value of $\Delta S_M$ occurs for temperatures and fields close to the multicritical point where these two peaks merge into a single peak. As the temperature is further increased this peak shifts to lower fields following the collinear AFM to paramagnet transition line, becomes broader and decreases its height.
Finally, at temperatures larger than $T_1(0)$ there is no peak in the MCE, only a monotonous reduction of the entropy with increasing magnetic field.

\subsection{Landau theory magnetocaloric effect}
The entropy change is obtained from the Landau functional using the thermodynamic relation $S_\text{M}=-\partial \mathcal{F}/\partial T$. Since the only explicit temperature dependence is in the parameters $A_Q$ and $A_0$, the entropy difference simply reads:
\begin{equation}
  \Delta S_\text{M}= %-\left.\frac{\partial{\mathcal{F}}}{{\partial T}}\right|_{H}+\left.\frac{\partial{\mathcal{F}}}{\partial T}\right|_{H=0}= 
  -A\left[s^2(B)+\frac{m^2(B)}{2}-s^2(0)-\frac{m^2(0)}{2}\right].
\end{equation}
The simplicity of this expression highlights the usefulness of the Landau functional approach to analyze de magnetocaloric effect. It allows to link the temperature and magnetic field behavior of the MCE to the corresponding one for the magnetization and the antiferromagnetic ordering.

In the paramagnetic phase, and for small $m$, the quartic term in $m$  can be neglected and minimizing  $\mathcal{F}_m$ we obtain $m=\frac{B}{A_0}$. This leads to a $-A \frac{m^2}{2}$ contribution to the entropy. A reduction of the magnetic entropy 
\begin{equation}\label{eq:paraMCE}
  \Delta S_\text{M}\sim -\frac{A B^2}{2 A_0^2}
\end{equation}
is therefore expected as the paramagnetic state is polarized with an external magnetic field.

\begin{figure}[t]
    \begin{center}
       \includegraphics[width=0.45\textwidth]{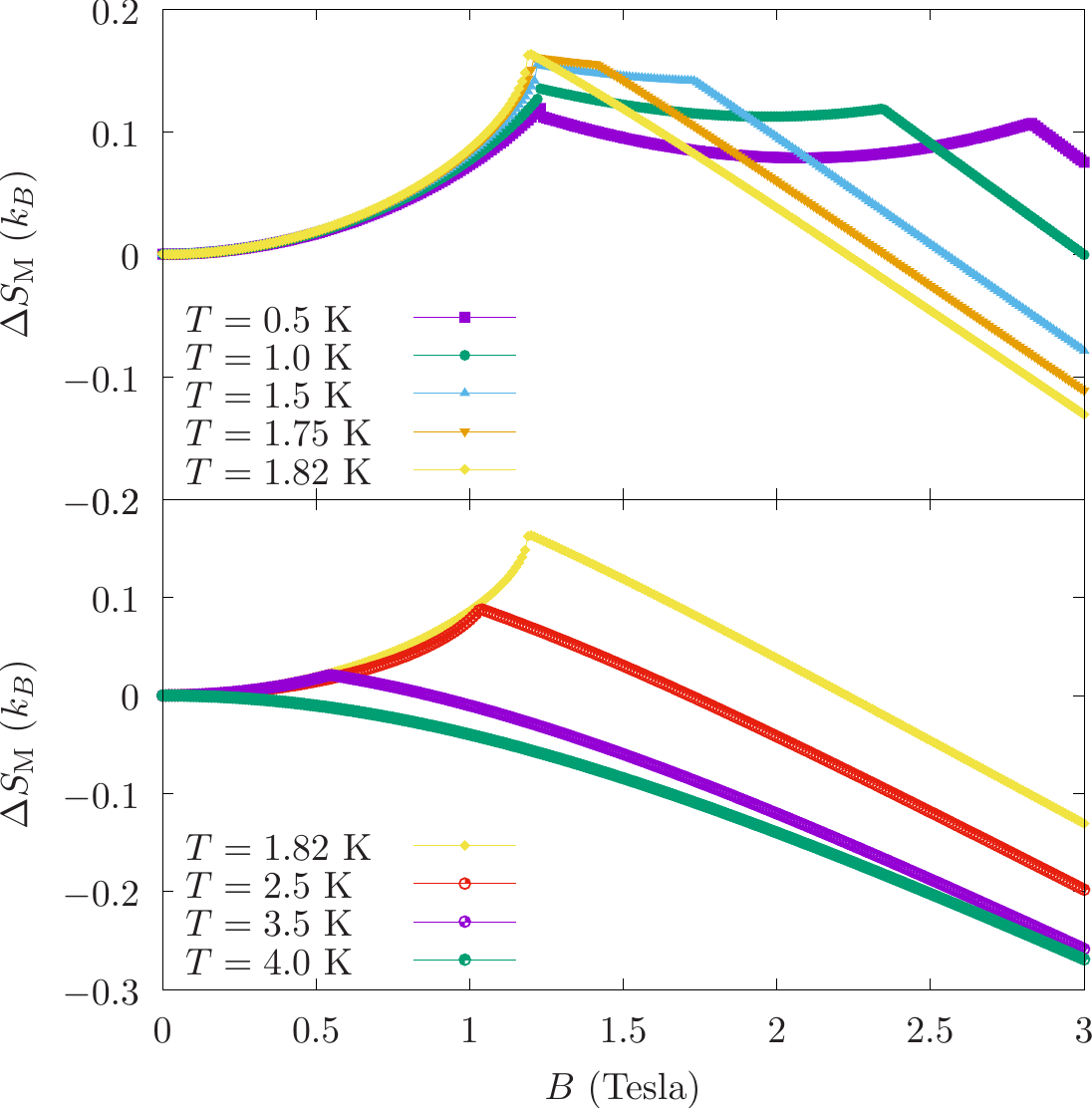}
    \end{center}
    \caption{Magnetic entropy change $\Delta S_\text{M}$ calculated using the Landau functional theory, for different values of the temperature and the final magnetic field. }
    \label{fig:cutslandau}
\end{figure}

To make a qualitative comparison with the Monte Carlo results of Fig. \ref{fig:MCCMC} we plot in Fig. \ref{fig:cutslandau} the MCE as a function of the final magnetic field for different values of the temperature. The proper units in the Landau functional are set multiplying it by $g\mu_B J\sim 4.69 k_B$ K/Tesla. The results obtained from the Landau functional reproduce the main features of the MCE obtained from the MC simulations. For temperatures below the multicritical point temperature we obtain a two peak structure with a wide plateau between them. The low and high temperature peaks are associated with the AFD to spin-flop and the spin-flop to paramagnetic phase transitions. The most significant differences with the Monte Carlo results are obtained at the lowest temperatures where the Landau functional underestimates the reduction of the MCE as the temperature is decreased. In general, the Landau functional is expected to perform best in the temperature and magnetic field regions where the order parameters are small. At temperatures of the order of the multicritical point temperature we obtain a single maximum where the highest value of the MCE is obtained. Increasing the temperature, the maximum decreases its intensity and shifts to lower magnetic fields, following the collinear AFM to paramagnetic transition line. For temperatures larger that $T_1(B=0)$ the behavior of the MCE is approximately quadratic in the field as expected from Eq. (\ref{eq:paraMCE}). 

\begin{figure}[t]
    \begin{center}
       \includegraphics[width=0.45\textwidth]{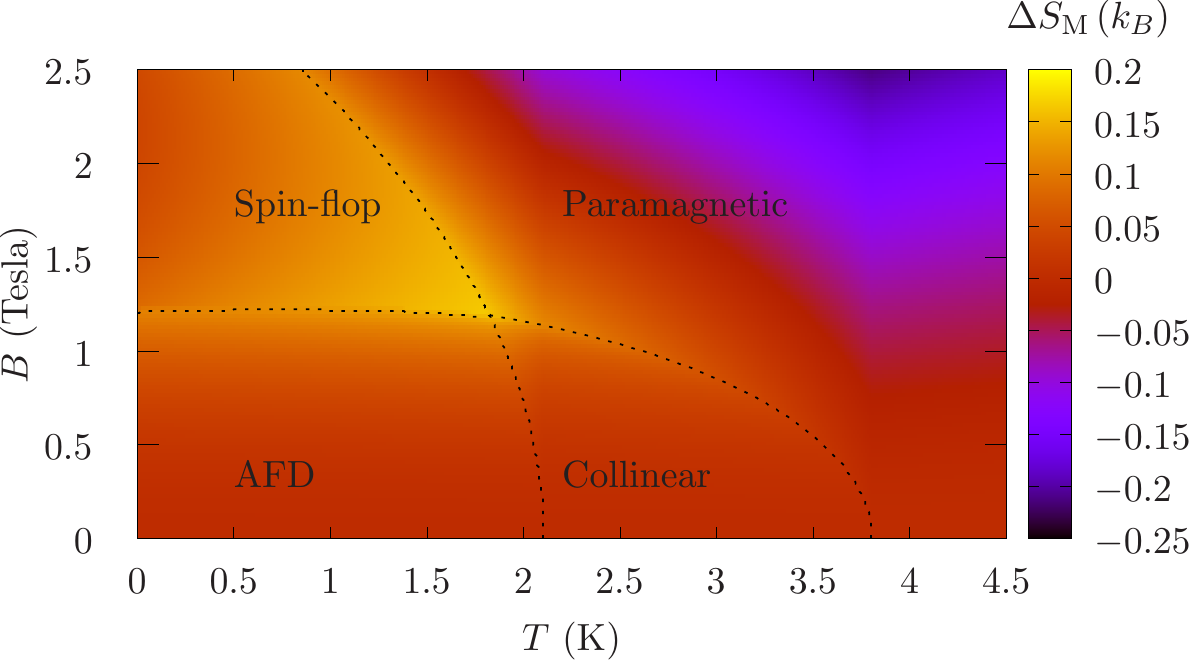}
    \end{center}
    \caption{Magnetic entropy change $\Delta S_\text{M}$ calculated using a Landau functional for different values of the temperature and the final magnetic field. }
    \label{fig:MClandau}
\end{figure}

The Landau functional results for $\Delta S_\text{M}$ in the $T$--$B$ plane are presented in Fig. \ref{fig:MClandau}. The spin-flop transition, and the antiferromagnetic to paramagnetic phase transitions can be clearly identified in the MCE data.   A large positive change in $\Delta S_\text{M}$ is observed for temperatures lower or of the order of the multicritical point temperature. The spin-flop transition is of the first-order type and there is a small jump in the entropy which is consistent with the Clausius-Clapeyron relation [Eq. (\ref{eq:CC})] and the small slope of the $B_f(T)$ curve. Interestingly, although the ideal collinear antiferromagnetic and AFD phases have a very different configurational entropy, only a weak anomaly is observed in the MCE data at the transition between these two phases. 

%\begin{equation}
%  \mathcal{F}_m=  \label{eq:fpara}
%\end{equation}
%Including a temperature dependence in $A_0$ as $A_0=A(T-T_=)$ to reproduce this behavior, 
%Here for simplicity we have chosen a temperature independent magnetic susceptibility, i.e a temperature independent $A_0$.

%The Curie-Weiss temperature

\section{Experimental results} \label{sec:EMCE}

\Eu~ single crystals (typical size: $\sim$ 0.70 $\times$ 0.10 $\times$ 0.10 mm$^3$) were obtained as described in Ref.\cite{franco2021synthesis} and oriented with a Laue machine. We found in all cases that the longest length of the needle shaped crystals corresponds to the crystallographic c direction. 
Magnetization data on the single crystals were obtained in a Quantum Design SQUID VSM magnetometer down to 1.8 K and up to 7 Tesla, with the applied field parallel to the c crystallographic direction.

\begin{figure}[t]
    \begin{center}
       \includegraphics[width=0.5\textwidth]{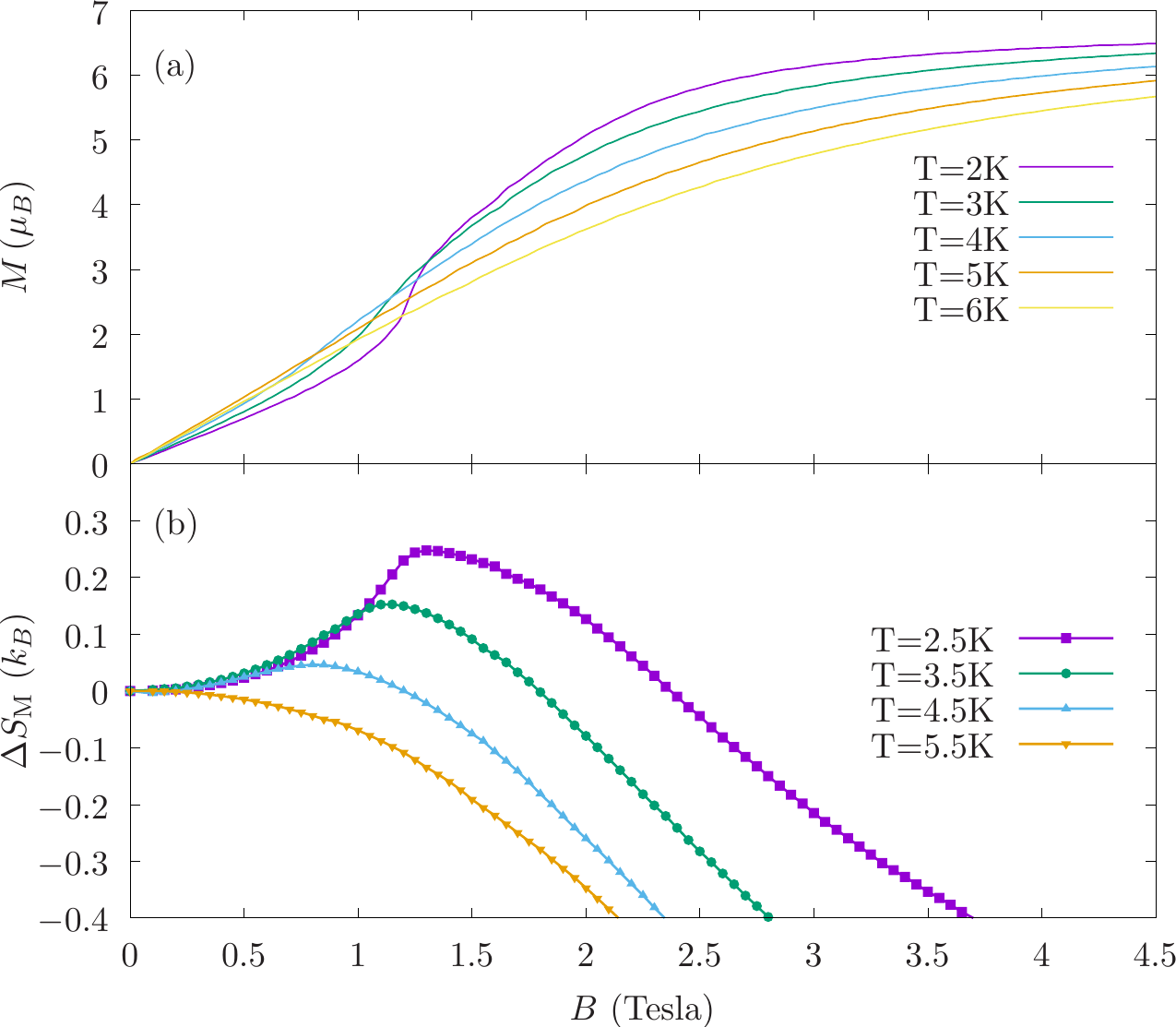}
    \end{center}
    \caption{(Color online) (a) Magnetization as a function of the external magnetic field ($B\parallel \hat{c}$) for a single crystal \Eu~sample. (b) Magnetic entropy change obtained from the magnetization data.}
    \label{fig:Figexp}
\end{figure}

The results for the magnetization as a function of the magnetic field for temperatures in the range (2K -- 6K). are presented in Fig. \ref{fig:Figexp} (a).
The kink observed in the lower temperature curves (T=2K,3K,4K) is consistent with the collinear AFM to PM transition expected from the theory. The kink shifts to lower magnetic fields as the temperature is increased and is suppressed for $T>5$K.
%The crystals were too small for a direct measurement of the specific heat.

The magnetic entropy change obtained using a discrete derivative (see Eq. (\ref{eq:maxwell})) of the experimental magnetization data is presented in Fig. \ref{fig:Figexp} (b). The highest temperature curve ($T=5.5$K) presents a monotonous decrease of the magnetic entropy with increasing magnetic field as expected in the paramagnetic phase. As the temperature is decreased, a peak with positive values of $\Delta S_M$ emerges which shifts to higher fields and increases its amplitude. At the lowest temperature attained, a broad peak in $\Delta S_M$ is observed with a maximum value of $\sim 0.25 k_B$ which is consistent with the maximum values expected from the theoretical analysis.  

\section{Summary and Conclusions}\label{sec:concl}

We have presented a detailed analysis of the magnetocaloric effect in \Eu\ using Monte Carlo simulations and a Landau functional approach. We based our Monte Carlo analysis on a model for the magnetic properties, with parameters estimated from {\it ab initio} calculations, that has been shown to provide a consistent description of the magnetic behavior observed experimetally for this compound.

We obtain a wide plateau of high magnetocaloric effect  $\Delta S_M \sim  5.14$ J kg$^{-1}$K$^{-1}$ for temperatures $T\lesssim 2.5$K and changes in the external magnetic field $\Delta B\lesssim 2$ Tesla. 
These values of $\Delta S_M$ are  comparable to the highest reported for other antiferromagnetic materials in the literature (see Ref. \onlinecite{midya2012giant} and references therein).
  %This analysis suggests that \Eu\ could be used for solid state refrigeration at low temperatures.

The Landau approach allowed us to provide a simplified description that captures the main features of the $T$ vs $B$ phase diagram, including three continuous phase transitions, a first-order transition and a multicritical point. It also provided a description of the MCE in terms of the behavior, as a function of the temperature and the external magnetic field, of the magnetization and an antiferromagnetic order parameter.

We also performed magnetization measurement experiments in  \Eu~single crystals that allowed us to obtain the magnetic entropy change for temperatures $T>2$K. The experimental results are in very good agreement with the expectations from the theory. Future work includes the measurement of this system at lower temperatures in order to explore the MEC in the full phase diagram and the growth of larger single crystals that would allow a direct measurement of the specific heat in a calorimeter.

Interestingly, although the magnetic frustration in the system leads to a rich phase diagram and to a reduction of the critical temperatures and fields, we found no clear indication that it is playing a role enhancing the MCE. In fact the collinear state is expected to present a much higher entropy, due to a high configurational degeneracy, than the AFD phase but the high MCE values obtained seem to be unrelated to this degeneracy. 

We expect our work to motivate further experimental studies on \Eu. In particular to obtain the adiabatic temperature change $\Delta T_{\text{ad}}$ for this material.
Our analysis may also be useful for the analysis of the MCE in ABX$_3$ compounds (A = alkali metal, B = transition metal, and X = halogen) which present a phase diagram similar to \Eu\cite{collins1997review,diep2013frustrated}. 

\acknowledgments
We thank Víctor F. Correa for useful discussions. We acknowledge financial support from ANPCyT grant PICT 2016/0204, SeCTyP-UNCuyo grant 06/C569.
\appendix
\section{Landau functional parameters} \label{ap:landau}

%The parameters used in the Landau functional of Sec. \ref{sec:landau} are presented in Table \ref{tab:LP}. 
We provide below analytical results for the transition lines, obtained from the minimization of the Landau functional, that were used to estimate the parameters in the Landau functional (see Table \ref{tab:LP}).

\subsection{Functional minimization and transition lines}

In the paramagnetic phase ($s=s_p=0)$ the magnetization can be obtained from the condition $\partial \mathcal{F}/\partial m=0$:
\begin{equation}
	A_0m+B_3m^3- B=0,
	\label{eq:para}
\end{equation}
which can be solved for the magnetization as a function of the temperature and the magnetic field.
\subsubsection{Paramagnetic to collinear AFM transition line}
The collinear AFM phase is characterized by $s_p=0$ and $\zeta=1$. The transition from paramagnetic to collinear AFM occurs when the coefficient of the $s^2$ term in $\mathcal{F}(\zeta=0,s_p=0)$ changes sign, i.e.:
\begin{equation}
A_Q-A_D+(B_5+2B_4)m^2+B_6 B^2=0.
	\label{eq:signchange_s2}
\end{equation}

In the absence of an external magnetic field the magnetization vanishes ($m=0$) and the transition occurs when $A_Q=A_D$ is satisfied. This leads to an expression for the paramagnetic to collinear AFM transition temperature: $T_1=T_Q+A_D/A$. 

In the general case, the transition line can be obtained solving Eqs. (\ref{eq:para}) and (\ref{eq:signchange_s2}). 
For $B_6=0$, the magnetization at the transition line can be readily obtained as a function of the temperature and the model coefficients:
\begin{equation}
	m=\sqrt{\frac{A_D-A_Q}{B_5+2B_4}}.
	\label{eq:mzb60}
\end{equation}

Replacing this expression for $m$ in Eq. (\ref{eq:para}) leads to a zero order (in $B_6$) expression for the critical field as a function of the temperature.
\begin{equation}
	B^0_{N1}=\sqrt{\frac{-A(T-T_1)}{B_5+2B_4}}\left( A_0-B_3\frac{A (T-T_1)}{B_5+2B_4} \right).
	\label{eq:H0N1}
\end{equation}

Including the lowest order correction in $B_6$ results in:
\begin{widetext}
\begin{equation}
	B^1_{N1}\sim B^0_{N1}\left( 1-B_6\frac{(-3A_1B_3+A_0(2 B_4+B_5))(-A_1B_3+A_0(2B_4+B_5))}{4(2B_4+B_5)^3} \right).
	\label{eq:HN1}
\end{equation}
\end{widetext}

\subsubsection{Collinear AFM to AFD transition line}

In the transition from collinear AFM to the AFD phase, $s_p$ becomes non-zero, which is associated with a change in the sign of the quadratic term in $s_p$ of the functional, i.e., with 
\begin{equation}
	A_D-2B_4 m^2 - 2 B_2 s^2=0.
	\label{eq:signchange_sp2}
\end{equation}
For $B=0$, we have $m=0$, and $\partial \mathcal{F}/\partial s=0$ leads to $s=\sqrt{\frac{A_Q-A_D}{B}}$ and the transition temperature $T_2=T_1-\frac{B_1}{2B_2}\frac{A_D}{A}$ can be readily obtained. 

For a finite magnetic field, $\partial \mathcal{F}/\partial s=0$ leads to:
\begin{equation}
  s^2=\frac{1}{B_1}\left(A_D-A_Q-\frac{1}{2}B_6 B^2-(2 B_4+B_5)m^2\right)
  \label{eq:dfds0}
\end{equation}
and $\partial \mathcal{F}/\partial m=0$ to:
\begin{equation}
  B_3 m^3+(A_0+2(2B_4+B_5)s^2)m-B=0.
  \label{eq:dfdm0}
\end{equation}
Replacing Eq.(\ref{eq:dfds0}) in Eq.(\ref{eq:dfdm0}) leads to a cubic equation in $m$. Once this equation is solved for $m$, the value of $s$  as a function of the temperature and the magnetic field can be obtained from Eq. (\ref{eq:dfds0}). The transition line is obtained using the values of $m$ and $s$ in Eq. (\ref{eq:signchange_sp2}).
\subsubsection{Paramagnetic to spin-flop phase transition line}

At the transition line from the spin-flop phase to the paramagnetic one: $\zeta=0$, $s=s_\perp \sqrt{2}$, and the coefficient of the quadratic term in $s$ in the functional changes sign, i.e.:
\begin{equation}
  A_Q+\frac{B_6B^2}{2}+B_5 m^2=0,
  \label{eq:floptopara}
\end{equation}
which, together with Eq. (\ref{eq:para}), can be solved for the critical magnetic field as a function of the temperature. For $B_5=0$ a simple analytical expression can be obtained:
\begin{equation}
  B_c=\sqrt{\frac{2 A (T_Q-T)}{B_6}}.
  \label{eq:floptoparaline}
\end{equation}
 
In our analysis we focused on a qualitative description of the Monte Carlo data and prioritized the simplicity of the functional.
Setting $B_2=B_1/2$, the expression for the transition temperature $T_2$ at zero field $T_2= T_1-\frac{B_1}{2B_2}\frac{A_D}{A}$ simplifies to  $T_2= T_1-\frac{A_D}{A}$. 
Using $T_1= T_Q+\frac{A_D}{A}$ and the results for the Monte Carlo transition temperatures at zero field we obtain the values for $T_Q$ and $A_D/A$.
The parameters $B_3$ and $B_4$ and $B_6$ set the value of the critical field $B_f(T)$ for the spin-flop transition and its temperature dependence. They were selected to obtain $B_f\sim 1.2T$ with little temperature dependence.
The parameter $B_5$ also modifies the slope of $B_f(T)$ and was set to zero for simplicity. %Finally, the parameter $T_0$ was chosen to be negative to avoid a spurious divergent behavior of the susceptibility in the antiferromagnetic phase.
%At high temperatures well in the paramagnetic state, the magnetic susceptibility is of the Curie-Weiss $\chi\propto 1/(T-\theta_{CW})$ type with
%\begin{equation}
%\theta_{CW} = \frac{J(J+1)}{3}(4 J_1+2 J_2)\sim 1\text{K}
%\end{equation}
%which has the same positive sign as in the experimental result.
\begin{table}[h]\label{tab:LP}
\begin{ruledtabular}
  \begin{tabular}{ll}
$A$& 1\\
$T_Q$& 2.1\\
$T_0$& -3.45\\
$A_D/A$ & 1.7\\
$B_1$ & 40\\
$B_2/B_1$ & $0.5$\\
$B_3/B_1$ & $0.3$\\
$B_4/B_1$ & $0.5$\\
$B_5/B_1$ & $0.0$\\
$B_6/B_1$ & $0.01$\\
\end{tabular}
\end{ruledtabular}
\caption{Parameters used in the Landau functional.}
\end{table}
To improve the accuracy of the fit to the Monte Carlo data, the functional could be extended to include a general temperature dependence of all the coefficients and higher order terms on the order parameters.

\bibliography{115b,chiral,triang,magnetocalorico}{}
\bibliographystyle{apsrev4-1}

\end{document}